\documentclass[pre,twocolumn,showpacs]{revtex4}

\usepackage{graphicx}
\usepackage{bm}

\newcommand{\avec}[1]{{\bm{#1}}}
\newcommand{\avecu}[1]{{\hat{\bm{#1}}}}

\begin{document}

\title{Simulated three-component granular segregation in a rotating drum}

\author{D. C. Rapaport}
\email{rapaport@mail.biu.ac.il}
\affiliation{Physics Department, Bar-Ilan University, Ramat-Gan 52900, Israel}

\date{June 06, 2007}

\begin{abstract}

Discrete particle simulations are used to model segregation in granular mixtures
of three different particle species in a horizontal rotating drum. Axial band
formation is observed, with medium-size particles tending to be located between
alternating bands of big and small particles. Partial radial segregation also
appears; it precedes the axial segregation and is characterized by an inner core
region richer in small particles. Axial bands are seen to merge during the long
simulation runs, leading to a coarsening of the band pattern; the relocation of
particles involved in one such merging event is examined. Overall, the behavior
is similar to experiment and represents a generalization of what occurs in the
simpler two-component mixture.

\end{abstract}


\maketitle

\section{Introduction}

One of the more fascinating properties of granular matter is the ability of
appropriately driven mixtures to separate into their individual components,
despite the apparent lack of energetic or entropic advantages of an unmixed
state. The segregation of a binary mixture contained in a partially filled,
horizontal, rotating drum is an extensively studied problem of this class, one
with obvious industrial importance. The most prominent characteristic exhibited
by this system is axial segregation, in which a pattern of bands of alternating
particle species forms along the cylinder axis; the bands merge and the pattern
coarsens over time, at a rate that drops to such a low level that it is unknown
whether the state eventually reached is stable, or merely longlived. A second
form of segregation, though one that can require greater effort to observe,
occurs in the radial direction, producing a core rich in small particles
surrounded by an outer layer of mainly big particles; this core can either be a
transient on the path to later axial segregation, or a feature that persists
even when full axial segregation is apparent from external views.

In the case of mixtures of two species, early experimental efforts involved
direct observation of the axial band structure at the outer surface
\cite{zik94}, whereas some of the subsequent studies succeeded in examining the
interior using MRI \cite{hil97f,nak97} and relating the development of the axial
bands to bulges occurring in the radial core. Time-dependent behavior can occur
subsequent to the initial appearance of the axial bands; narrower bands
gradually merge to form broader bands \cite{fre97} and traveling surface waves
have been noted \cite{cho98}. Complicating the behavior are results showing that
the ratio of cylinder to particle diameter determines if axial segregation can
occur and whether its appearance depends reversibly on rotation rate
\cite{ale04}. Further recent experimental examples \cite{arn05,fin06} reveal
more about the richness of the segregation effect and the associated dynamics
under wet conditions; these are relevant because the behavior of slurries tends
to be similar to dry mixtures. Overall, no consensus has yet emerged as to the
nature of the mechanisms underlying this complex set of segregation phenomena.
For further discussion of the difficulties inherent in understanding granular
matter in general see, e.g., Refs.~\cite{jae96,kad99,ara06}.

In contrast to the many studies involving two-component systems (only a
selection of which were cited above), three-component systems have received far
less attention. The experimental results available for systems with three
particle species \cite{new04} were obtained using glass beads of different sizes
(an earlier brief discussion appears in \cite{das91}). The description of the
experiment relates that, after beginning with an initially uniform mixture, the
first effect to occur is the appearance of radial segregation, in which there is
an inner core of small particles and an outer layer of big particles, with the
medium-size particles lying predominantly in between. This is followed by the
development of axial bands of medium-size particles in the mainly big-particle
outer region, followed by the appearance of small-particle bands within the
medium-particle bands. Replacing the opaque big and/or medium particles with
transparent equivalents reveals that the small-particle core persists after the
onset of axial segregation, and that a layer of medium-size particles surrounds
it. Systems with more than three components are also mentioned, with four
components showing behavior analogous to three, while beyond four the axial
segregation is no longer seen, but these lie outside the scope of the present
work. As with two-component systems, the behavior is certain to be influenced by
the relative species sizes and concentrations, as well as by the many parameters
required to fully specify the system, including the fill level, frictional
properties and rotation speed, although their influence on the nature of the
segregation is not described.

Simulations of granular systems based on discrete-particle models have proved
capable of reproducing both the axial and radial forms of segregation, in
qualitative agreement with experiment. The present paper extends an earlier
study \cite{rap07}, in which two-component systems were modeled. In that paper
it was shown how the appearance of radial segregation preceded axial
segregation, and that depending on the circumstances, a radial core of small
particles could still be present even after the exterior view showed essentially
complete axial segregation. Owing to the need for long runs and large systems,
only a limited set of parameter combinations could be considered. The problem of
following the development of segregation in detail is further exacerbated by a
lack of reproducibility between runs differing only in the random velocities
assigned to the particles in the initial state that result in different
segregation scenarios. While similar effects occur experimentally, the
implication is that multiple simulation runs are essential to ensure that the
`typical' behavior is captured. In view of the potentially heavy computational
demands, a thorough study of the two-component system and a detailed comparison
with the variety of experimental data available has yet to be undertaken.

The obvious next stage in this exploration is to increase the number of granular
species, and while this has been carried out experimentally, as indicated above,
the corresponding simulations have yet to be carried out. The goal of the
present paper, therefore, is to demonstrate that the model used previously is
also able to produce the appropriate kinds of segregation even when three
species are involved. Given the very limited amount of experimental information
presently available for this problem, any systematic coverage of the parameter
space is premature; the discussion is therefore confined to specific examples
illustrating the main features of simulated three-component segregation.

\section{Methodology}

The granular model employs spherical particles whose overlap repulsion is based
on a continuous potential \cite{cun79,wal83,haf86}, exactly as in \cite{rap07}.
The normal force between a pair of particles $(i, j)$ includes a repulsion that
depends linearly on the overlap and a velocity-dependent damping force,
\begin{equation}
\avec{f}_n = \left[ k_n ( d_{ij} - r_{ij} ) - \gamma_n (\avecu{r}_{ij} \cdot
\avec{v}_{ij}) \right] \avecu{r}_{ij} , \quad r_{ij} < d_{ij}, \label{eq:fn}
\end{equation}
where $\avec{r}_{ij}$ and $\avec{v}_{ij}$ are the particle separation and
relative velocity, and $d_{ij} = (d_i + d_j) / 2$ is the mean diameter. The two
parameters appearing in Eq.~(\ref{eq:fn}), $k_n$ governing the stiffness and
$\gamma_n$ the normal damping, are the same for all particles.

Sliding of particles that are within interaction range is opposed by both
frictional damping and a transverse restoring force,
\begin{equation}
\avec{f}_t = - \gamma_s^{c_i c_j} \avec{v}_{ij}^s -
 k_g \int_{\rm (coll)} \avec{v}_{ij}^s (\tau) \, d \tau , \label{eq:ft}
\end{equation}
where $\avec{v}_{ij}^s$ is the relative transverse velocity allowing for
particle rotation, and the integral is evaluated as the sum of vector
displacements over the collision period (additionally, whenever the integrated
displacement exceeds 0.1 it is reset to zero). The magnitudes of both terms in
Eq.~(\ref{eq:ft}) are bounded by $\mu^{c_i c_j} |\avec{f}_n|$. The transverse
damping and static friction coefficients, $\gamma_s^{c_i c_j}$ and $\mu^{c_i
c_j}$, depend on the particle types $c_i$ and $c_j$; the transverse stiffness
$k_g$ is the same for all particles.

The majority of parameter settings are taken from \cite{rap07} and, as before,
the curved cylinder wall and the flat end plates are treated as rough and smooth
boundaries, respectively. The value of the gravitational acceleration, $g = 5$,
relates the dimensionless MD (molecular dynamics) units of the simulation to the
corresponding physical units. If $L_{MD}$ is the length unit (in mm), then the
time unit is $T_{MD} \approx 10^{-2} \sqrt{5 L_{MD}}$\,s. A cylinder rotating
with angular velocity $\Omega$ (MD units) has an actual rotation rate of $\Omega
/ (2 \pi T_{MD}) \approx 7.1 \Omega / \sqrt{L_{MD}}$\,Hz; for 4\,mm particles,
$\Omega = 0.2$ is equivalent to an experimental 43\,rpm. Other details of the
simulations are covered in \cite{rap02}, and the required MD techniques are
described in \cite{rap04}.

Efficiently evaluating the displacement sums in Eq.~(\ref{eq:ft}) requires
maintaining a list of all currently interacting particle pairs in a form that is
readily accessible given the identity of the pair involved, but also easily
modified as pairs are added or deleted due to the continually changing
identities of interacting neighbors. Of the approaches available for organizing
the data, the one used here (as well as in \cite{rap07}, although the
description was omitted) is the following.

A table is constructed that holds a set of linked lists of interacting pairs
$(i, j)$, ordered so that $i < j$. Each entry consists of the value of $j$, a
pointer to another entry used to link entries with a common $i$, the current
value of $\avec{u}_{ij} = \sum_{\rm (coll)} \avec{v}_{ij}^s (\tau)$ needed in
Eq.~(\ref{eq:ft}), and the timestep number $t_{ij}$ at which $\avec{u}_{ij}$ was
last updated. There is a separate list for each particle $i$, and a pointer
$h_i$ to the first entry in its list; from $h_i$ the list can be scanned using
the pointers to obtain all interacting neighbors with $j > i$, until a null
pointer ends the list. For each interacting pair encountered during the force
calculations, the corresponding $\avec{u}_{ij}$ and $t_{ij}$ are updated; new
entries are readily added as necessary, whereas if a previously interacting pair
is no longer in range its entry will not be updated. After all the forces have
been evaluated, those entries whose $t_{ij}$ are not current (because $r_{ij} >
d_{ij}$) have their $\avec{u}_{ij}$ reset to zero. Expired entries need not be
deleted immediately since pairs are capable of moving in and out of range many
times, in which case, if the relevant entry is still in the table it will be
reused. (On a shared-memory computer, if the neighbor list references the
relevant table entries, the force computations can be carried out in parallel
\cite{rap04}, even with this history-dependent interaction.)

The force parameters are $k_n = 1000$, $\gamma_n = 5$, and $k_g = 500$.
Transverse damping (or dynamic friction) depends on particle type; if $b$, $m$
and $s$ denote big, medium and small sizes, then $\gamma_s^{bb} = 10$,
$\gamma_s^{mm} = 6$, $\gamma_s^{ss} = 2$, and the lower of the values is used
for unlike species (the $bb$ and $ss$ values are from \cite{rap07}, and the $mm$
value is their mean). Finally, the relative values of the static friction
coefficients, e.g., $\mu^{bb} / \mu^{ss}$, are set equal to the ratio of the
corresponding $\gamma_s$ values, with the larger of the pair equal to 0.5. In
general, there seems to be considerable latitude in choosing the parameter
settings, and what is required at this exploratory stage is the ability to
produce results that are qualitatively reasonable.

The nominal particle diameters correspond to the interaction cutoff. As in
\cite{rap07}, for small particles $d_s = 2^{1/6} \approx 1.122$ (MD units). For
big particles $d_b = 2 d_s$, and for medium particles $d_m = 1.3 d_s$; this
choice of relative sizes is not too different from that used experimentally. The
actual small-particle diameters are uniformly distributed over a narrow range $[
d_s - 0.2, d_s ]$ (the mean diameter of the small particles is then close to
unity), and likewise for the other species. Particles all have the same material
density.

\section{Results}

Three simulation runs are singled out for analysis here, each aimed at
demonstrating particular features of the behavior. They employ cylinders with
diameter $D = 30$ or $40$ (reduced units) that are rotated with angular velocity
$\omega = 0.2$ where particle flow is continuous; these are the values used for
the majority of runs in the two-component study \cite{rap07}. Runs are begun
with the particles arranged on a lattice at a specified density (this determines
the eventual fill level) and assigned random velocities. Particle species is
assigned randomly, producing an initially mixed state; the relative populations
of the three kinds of particles are chosen to ensure equal volume fractions of
all species.

The first of the runs, \#{\em A}, involves a cylinder of length $L = 240$ and
diameter $D = 30$ (aspect ratio $L / D = 8$); the initial filling density is
$\rho = 0.25$, resulting in a system with $N = 29\,704$ particles. The total run
length is $n_R = 4260$ revolutions, and the purpose of this comparatively long
run (by simulation standards) is to demonstrate the ability to produce an
apparently stable axial band pattern after the effects of the initial transients
have disappeared. The second run, \#{\em B}, employs a longer cylinder, $L =
360$, with the same $D$ and filling density, so that $N = 44\,556$; the run
length $n_R = 2300$ is shorter, and the higher aspect ratio ($= 12$) can
accommodate a more extensive set of axial bands. In the third run, \#{\em C},
the cylinder length is reduced to $L = 120$, the diameter increased to $D = 40$,
and the fill level raised by setting the initial density to $\rho = 0.4$ so that
$N = 47\,520$; this is a shorter run whose goal is to provide increased space
for observing the development of radial segregation.

The behavior encountered in each of these runs is described in detail below.
This is accomplished using a combination of axial space-time plots that
summarize the overall development of axial segregation, graphs showing the
concentrations of each of the particle species along the axial and radial
directions, and images of individual configurations that reveal where particles
actually reside.

\begin{figure}
\includegraphics[scale=1.13]{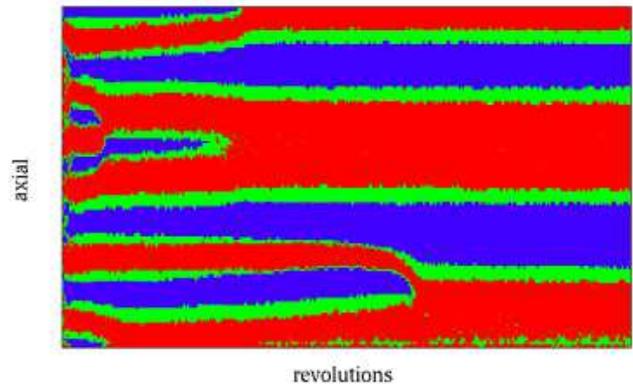}
\caption{\label{fig:f01} (Color online) Axial space-time plot for run \#{\em A}
($L = 240$, $L / D = 8$, $n_R = 4260$); red, green and blue (or medium, light
and dark shades of gray) denote whether the big, medium or small particles have
the highest volume fraction. Time is along the horizontal axis, and the vertical
axis corresponds to axial position.}
\end{figure}

Figure~\ref{fig:f01} shows the axial space-time plot for run \#{\em A}. In view
of the need to distinguish between three species, rather than just two, the
color coding shows only the volume-weighted majority species, irrespective of
the degree to which the other species might be present; in this respect the
present space-time plots differ from \cite{rap07} where a continuous color
gradient designates the relative volume fraction itself. The plot shows that
during slightly over half of the run a series of band merging events occurs,
most within the first 1200 revolutions and the last at approximately 2500,
eventually producing a state with 10 axial bands (counting all species, where
one of the end bands is only weakly defined), and there is no apparent change
thereafter. While the possibility of future change cannot be excluded, none was
observed when the run was subsequently extended by an extra 1000 revolutions
(not shown). The rapid, small fluctuations at the band boundaries are artifacts
of the discrete nature of the majority rule by which color is determined.

\begin{figure}
\includegraphics[scale=0.85]{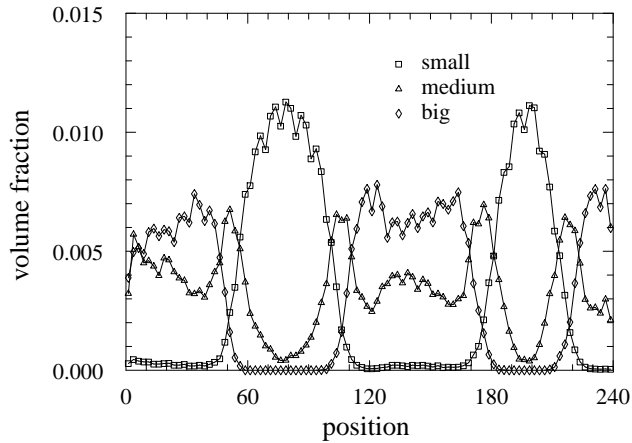}
\caption{\label{fig:f02} Volume fractions of the three particle species measured
in slices normal to the axis for the axially segregated state near the end of
run \#{\em A} (distance is expressed in dimensionless MD units).}
\end{figure}

More detailed information concerning the band occupancy at the end of run \#{\em
A} appears in Fig.~\ref{fig:f02}. The plot shows the volume fractions of each of
the particle species in a sequence of slices oriented normal to the cylinder
axis; fluctuations in the measurements are reduced by averaging over ten
successive configurations, sampled once per revolution. It is clear that the $b$
and $s$ particles are well segregated axially. On the other hand, the axial
separation of $m$ and $b$ particles is less pronounced than for $m$ and $s$;
while the $m$ particles exhibit a distinct preference for the regions located
between $b$ and $s$ bands, and their occupancy drops to a very low level in the
middle of the $s$ bands, they are seen to maintain a significant minority
presence throughout the $b$ bands.

\begin{figure}
\includegraphics[scale=0.65]{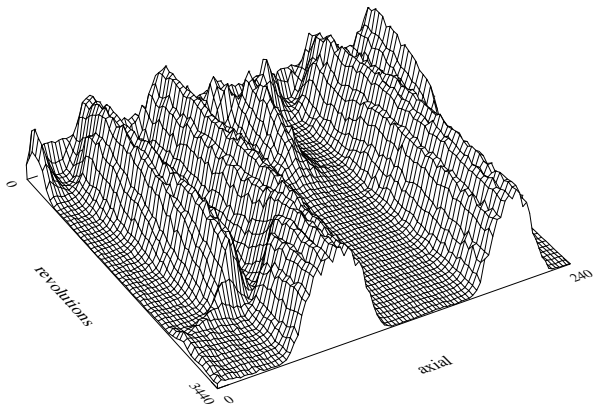}
\caption{\label{fig:f03} Development of the peaks in the the axial distribution
of small particles for run \#{\em A}.}
\end{figure}

Adding an extra time dimension to Fig.~\ref{fig:f02} provides a more detailed
description of the segregation process than is available in the space-time plot,
but the results for the three species must be graphed separately to ensure the
features are visible. The time-dependent distribution of small particles is
displayed as a surface plot in Fig.~\ref{fig:f03}; prominent features include
the emergence of the peaks in the axial distribution and the subsequent peak
changes corresponding to axial band reorganization.

\begin{figure}
\includegraphics[scale=0.85]{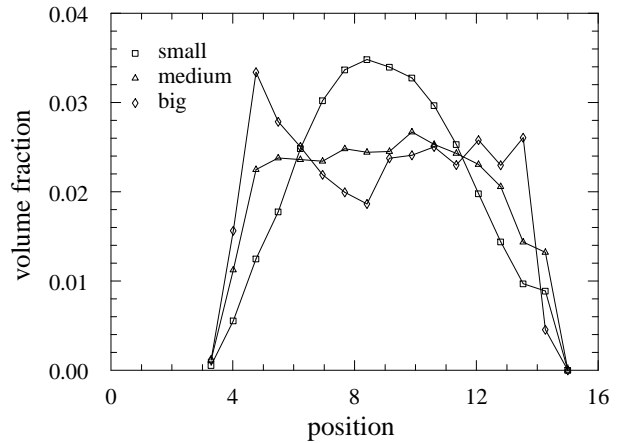}
\caption{\label{fig:f04} Volume fractions of the particle species early in run
\#{\em A} measured in slices parallel to the free surface (the cylinder axis is
at the origin).}
\end{figure}

Figure~\ref{fig:f04} shows the volume fractions at an early stage of run \#{\em
A} (after approximately 37 revolutions, prior to the appearance of any axial
effects) in a manner aimed at revealing the presence of radial segregation. In
order to achieve this, the system is divided into a series of slices that are
oriented parallel to the mean direction of the upper free surface of the mixture
(as in \cite{rap07}, the slope is determined by a linear fit to the inner 2/3 of
the surface away from the curved boundary) and the relative volume fractions in
the slices evaluated (also limited to a region of the same width to avoid bias);
the results are again averaged over ten configurations, at one-revolution
intervals, for improved statistics. The $b$ and $s$ particles are seen to
experience a certain degree of radial segregation, with $s$ dominating the
interior and $b$ on the outside, but the $m$ particles exhibit no radial
preference (the behavior is slightly different in run \#{\em C}, below, where
the particle layer has double the maximum thickness due to larger $D$ and a
higher fill level).

\begin{figure}
\includegraphics[scale=1.13]{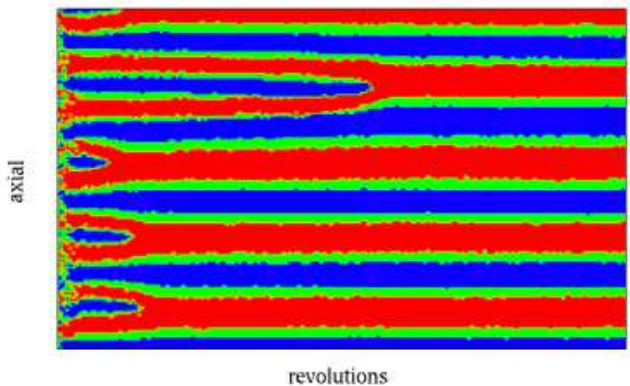}
\caption{\label{fig:f05} (Color online) Axial space-time plot for run \#{\em B}
($L = 360$, $L / D = 12$, $n_R = 2300$).}
\end{figure}

\begin{figure}
\includegraphics[scale=0.85]{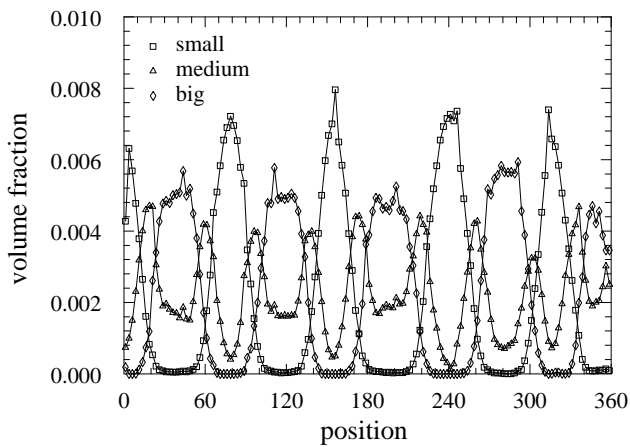}
\caption{\label{fig:f06} Volume fractions of the particle species in the axially
segregated state near the end of run \#{\em B}.}
\end{figure}

Figure~\ref{fig:f05} shows the axial space-time plot for run \#{\em B}, where
the longer cylinder allows the formation of a greater number of axial bands.
Several band merging events are apparent, most during the first 400 revolutions
and the last at about 1200, resulting in 19 bands at the end of the run; the
final band merge will be examined in greater detail below. A plot of the volume
fractions as a function of axial position at the end of the run is shown in
Fig.~\ref{fig:f06}; once again the $b$ and $s$ particles are well segregated,
while the $m$ particles exhibit the same preference for the regions between the
$b$ and $s$ bands and a minority presence inside the $b$ bands, as in run \#{\em
A}.

\begin{figure}
\includegraphics[scale=0.72]{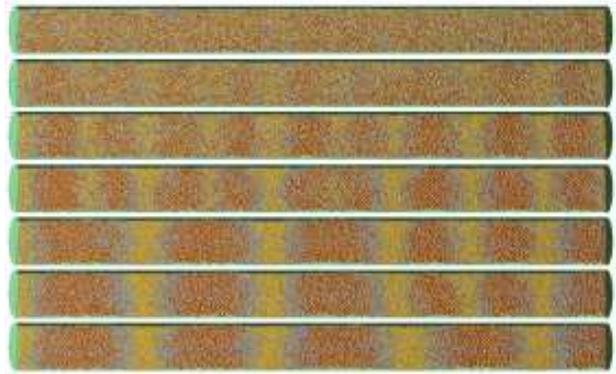}
\caption{\label{fig:f07} (Color online) Surface views of run \#{\em B} at
different times.}
\end{figure}

While the colored space-time plots only allocate a single column of pixels to
describe the state of the entire system at each instant, images of the evolving
patterns viewed from above provide a more detailed history resembling what would
be seen experimentally. Figure~\ref{fig:f07} shows the state of run \#{\em B}
after approximately 20, 60, 100, 200, 400, 800 and 1600 revolutions; the images
cover the early portion of the run where there is no hint of axial segregation
(although there are radial processes occurring beneath the surface), followed by
the pattern coarsening stage, and the steady state that persists throughout the
latter part of the run.

\begin{figure}
\includegraphics[scale=1.20]{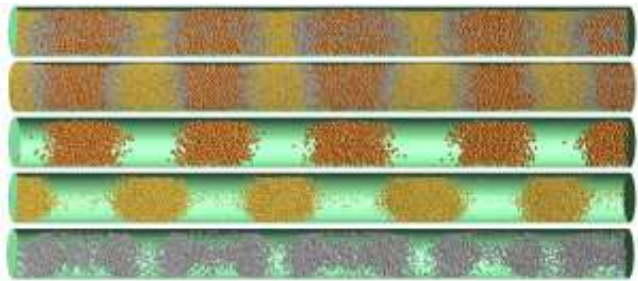}
\caption{\label{fig:f08} (Color online) Final state of run \#{\em B}; the full
system seen from above and below, and separate views of the big, small and
medium size particles, colored copper, gold and silver, respectively).}
\end{figure}

Figure~\ref{fig:f08} contains views of the final state of run \#{\em B} as seen
from the outside, along with images of the individual species that correspond to
what might be observed experimentally, either using MRI, or by replacing any two
of the species with transparent particles having similar mechanical properties.
The first two views are of the system seen from above and below (the view
direction is approximately normal to the free surface); the remaining views each
show a single particle species. The $b$ and $s$ particles are seen to occupy
well-defined axial bands; the $m$ particles, however, show only partial axial
segregation, an observation confirmed quantitatively in Fig.~\ref{fig:f06}. A
feature apparent from the pictures is that the boundaries of the $b$ and $s$
bands are not vertical, with the $s$ bands being narrower at the upper free
surface.

\begin{figure}
\includegraphics[scale=0.85]{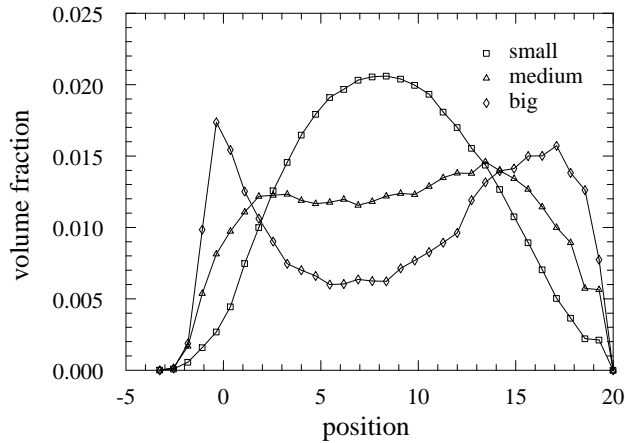}
\caption{\label{fig:f09} Volume fractions of each particle species early in run
\#{\em C} measured in slices parallel to the free surface (the cylinder axis is
at the origin).}
\end{figure}

\begin{figure}
\includegraphics[scale=1.43]{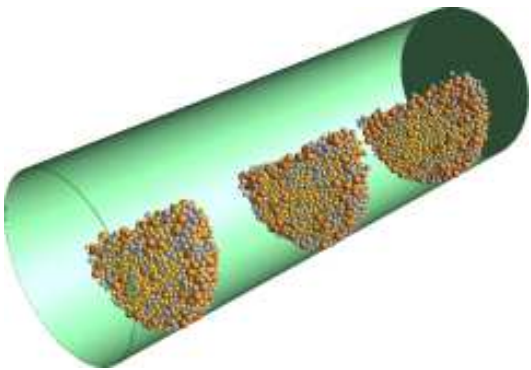}
\caption{\label{fig:f10} (Color online) Run \#{\em C} after 35 revolutions;
three narrow slices through the system are shown.}
\end{figure}

The distribution of volume fractions for run \#{\em C} after 35 revolutions,
evaluated as before to demonstrate the presence of radial segregation (also
averaged over ten configurations), is shown in Fig.~\ref{fig:f09}. Unlike the
case of run \#{\em A} earlier, the behavior is more clearly resolved due to the
larger cylinder diameter and higher fill level. There is now also a suggestion
of the $m$ particles having preferred radial positions, although the effect is
still much weaker than for the $b$ and $s$ particles. A view showing three
slices cut out of the system appears in Fig.~\ref{fig:f10}; what can be seen
here reinforces the conclusion from Fig.~\ref{fig:f09}, that despite the noise,
the $b$ and $s$ particles show distinct preferences for the exterior and
interior, respectively, with the $m$ particles also preferentially located
towards the exterior, though to a lesser degree than $b$ (axial segregation
begins later in this run, although the low aspect ratio impedes its
development).

When comparing the simulation results with experiment, apart from any possible
shortcomings of the model, allowance must be made for the relatively small
system sizes that can affect the results. Given that the computational
requirements depend on the product of $N$ and $n_R$, the need to avoid
excessively long simulations means that the ratio of cylinder diameter to
particle size, $D / d_s$, must be smaller than in experiment; in addition to
finite-size effects in general, this particular limitation has been shown to
influence behavior experimentally \cite{ale04}. Nevertheless, to the extent that
it is possible to compare the three-component simulations with the limited
experimental data available \cite{new04}, the similarity is encouraging, in
particular, the organization of the axial bands, with the bands of $m$ particles
located between alternating $b$ and $s$ bands. The fact that experiment
indicates that a core of $m$ particles persists within the $b$ bands is also in
agreement with the simulations, even if in the latter the core is less well
defined.

As always, however, it is the deviations from experiment that must be
considered. For example, experiment suggests that, in addition to the axial $m$
bands being more sharply defined, there is apparently an innermost core of $s$
particles (surrounded by $m$ particles). Whether such differences, in particular
the latter, are due to basic flaws in the model, or whether they are merely
consequences of a less than ideal parameter choice (perhaps as simple as the
range of particle sizes), or too narrow a cylinder, must await future study,
both simulational and experimental.

\begin{figure}
\includegraphics[scale=1.40]{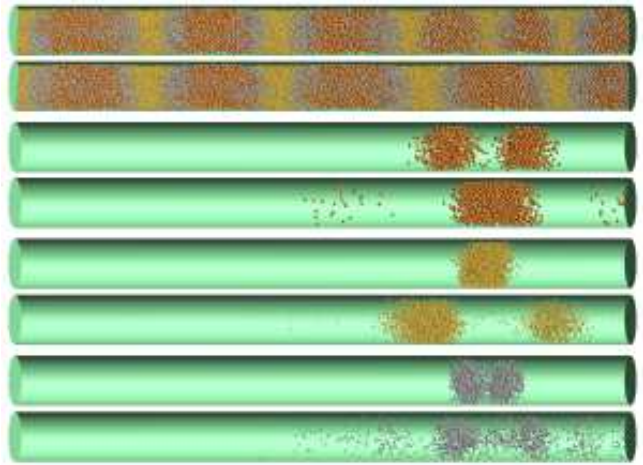}
\caption{\label{fig:f11} (Color online) Band merging in run \#{\em B}; the views
show the full system and the individual particles from the $b$, $s$ and $m$
bands involved (only the inner portions of the $m$ bands), both before and after
the event.}
\end{figure}

The final example considered involves the band merging event of run \#{\em B}
that occurs between revolutions 700 and 1450 at a location in the cylinder
sufficiently far from the end plates to be free from any interference. An
analogous study was carried out for the two-species system \cite{rap07}, but
with three species the details are more complex. Figure~\ref{fig:f11} shows what
becomes of the occupants of the merging $b$ bands and the vanishing $s$ and $m$
bands. The first two images are of the entire system seen from above, before and
after the event; pattern changes are limited to the neighborhood of the merging
bands. Each subsequent pair of images shows the particles of a single species
within the affected bands, both before the event (including particles beneath
the surface intruding into adjacent bands) and afterwards; in the case of the
$m$ particles, only those initially in the inner halves of the two bands are
considered. The $b$ bands are seen to merge with only the slightest scatter
outside the newly formed band. The $s$ band disperses into the adjacent bands
with practically no scatter beyond the bands receiving these particles. In the
case of the $m$ bands (where only the inner half-bands are tracked) the contents
are seen to disperse to a greater degree than the $b$ bands; such behavior is
consistent with the fact that axial segregation, and hence confinement, is
weakest for the $m$ particles. 

\section{Conclusion}

Particle-based simulations of three-component granular systems in a rotating
drum have been used in demonstrating that both axial and radial segregation can
be produced using a simple model for the granular particles and their
interactions. In the case of the two-component systems considered previously,
there was already a relatively large set of parameters whose settings are
potentially capable of influencing the behavior; with three components the set
is even larger. Exploring the consequences of systematic parameter variations on
the behavior, and developing a closer connection between simulation and
experiment, are subjects for future study.

\bibliography{threecomp}

\end{document}